\documentclass[12pt, a4paper]{article} 

\usepackage[includeheadfoot, width=442pt, height=720pt]{geometry}
\usepackage{amsmath}
\usepackage{amsfonts}
\usepackage{amssymb}
\usepackage{bbm,bm}
\usepackage{mathtools}
\usepackage{slashed}
\usepackage{mathalfa}
\usepackage{graphicx,caption,subfigure}
\usepackage{tikz} 
\usepackage{tikz-cd}
\usetikzlibrary{decorations.pathreplacing,calc,tikzmark}
\usepackage{booktabs}
\usepackage{adjustbox}
\usepackage[utf8]{inputenc}
 \usepackage[splitrule]{footmisc}

\usepackage{placeins}
\usepackage{empheq}
\usepackage{paralist}
\usepackage{cite}
\usepackage[normalem]{ulem}
\usepackage{soul}

\usepackage[colorlinks=false,urlbordercolor=red]{hyperref}
\usepackage{xcolor}
\usepackage{tensor}

\newcommand{\nc}{\newcommand}
\nc{\lb}{\llbracket}
\nc{\rb}{\rrbracket}
\nc{\gl}{\llbracket}
\nc{\gr}{\rrbracket}
\nc{\del}{\partial}
\nc{\tri}{\hspace{-3.5pt}\vartriangle\hspace{-3.5pt}}
\nc{\blacktri}{\blacktriangle}

\nc{\eq}[1]{\begin{equation}
                     \begin{split} #1 \end{split}
                     \end{equation}}

\nc{\ov}{\overline}

\nc{\fa}{\hat}
\nc{\fb}{\MakeUppercase}
\nc{\fc}{\tilde }
\nc{\Lie}{{\cal L}} 
\nc{\lambdabar}{{\mkern0.75mu\mathchar '26\mkern -9.75mu\lambda}}

\allowdisplaybreaks[2]

\DeclareUnicodeCharacter{2212}{-}
\begin{document}

\vspace*{-1.5cm}
\begin{flushright}
  {\small
  LMU-ASC 56/22\\
  MPP-2022-289
  }
\end{flushright}

\vspace{1.5cm}
\begin{center}
  {\Large \bf Black hole entropy and moduli-dependent\\[.3cm] species scale} 
\vspace{0.65cm}

\end{center}

\begin{center}
{\large
Niccol\`o Cribiori$^a$, Dieter L\"ust$^{a,b}$ and Georgina Staudt$^{a,b}$
}
\end{center}

\vspace{0.1cm}
\begin{center} 
\emph{
$^a$Max-Planck-Institut f\"ur Physik (Werner-Heisenberg-Institut), \\[.1cm] 
   F\"ohringer Ring 6,  80805 M\"unchen, Germany, 
   \\[0.1cm] 
 \vspace{0.3cm}
$^b$ Arnold-Sommerfeld-Center for Theoretical Physics,\\ Ludwig-Maximilians-Universit\"at, 80333 M\"unchen, Germany \\[.1cm] 
    } 
\end{center} 

\vspace{0.5cm}

\begin{abstract}

We provide a moduli-dependent definition of species scale in quantum gravity based on black hole arguments. 
Concretely, it is derived from a lower bound on the entropy of extremal black holes with higher curvature corrections, which ensures that the black hole can be reliably described within the effective theory.
By demanding that our definition coincides with a recent proposal for a moduli-dependent species scale motivated from the topological string, we conclude that the conjecture $\mathcal{Z}_{BH} = |\mathcal{Z}_{\rm top}|^2$ relating the black hole to the topological string partition functions should hold, at least within the regime of validity of our analysis.

\end{abstract}

\thispagestyle{empty}
\clearpage

\setcounter{tocdepth}{2}

\section{Introduction}

Effective theories are characterized by at least two energy scales: an ultraviolet cutoff $\Lambda_{UV}$ and an infrared cutoff $\Lambda_{IR}$. 
They define the regime of validity of the description. 
The focus of the present letter is on the scale $\Lambda_{UV}$ in effective theories of gravity and especially in $d=4$ spacetime dimensions.

While a natural candidate for the UV cutoff in this context is the Planck scale $M_P$, it has been argued \cite{Dvali:2007hz,Dvali:2007wp,Dvali:2010vm,Arkani-Hamed:2005zuc,Distler:2005hi,Dimopoulos:2005ac,Dvali:2009ks,Dvali:2012uq} that the presence of a number $N$ of light degrees of freedom lowers the expected UV cutoff down to a scale, $\Lambda_{\rm sp} \simeq {M_P}{N^{1/(2-d)}}$, called species scale. 
In the spirit of the swampland program \cite{Palti:2019pca, Agmon:2022thq}, one can say that the naive effective theory expectation is altered by non-trivial (quantum) gravity effects. 
Indeed, the role of the species scale in swampland conjectures has received attention e.g.~in \cite{Cribiori:2021gbf, Castellano:2021mmx,Lust:2022lfc, Long:2021jlv,Montero:2022prj,Blumenhagen:2022dbo,Blumenhagen:2022zzw,Castellano:2022bvr}.

In the recent work \cite{vandeHeisteeg:2022btw}, a moduli-dependent definition of species scale has been proposed. The motivation stems from the fact that, at least within the regime of computational control, supergravity effective theories typically contain a number of massless scalar fields. These parametrize the relevant couplings and scales of the theory, and $\Lambda_{\rm sp}$ should be no exception. The proposal of \cite{vandeHeisteeg:2022btw} captures the moduli dependence of $\Lambda_{\rm sp}$ in the context of $\mathcal{N}=2$ supergravity theories in four dimensions and it is checked in asymptotic limits in the moduli space in several examples.

The argument of \cite{vandeHeisteeg:2022btw} exploits properties of type II string theories compactified on a Calabi-Yau threefold and relates the species scale to the genus-one free energy $F_1$ of the topological string propagating on the mirror-dual \cite{Bershadsky:1993ta, Bershadsky:1993cx}.
As a result, the number $N$ of light degrees of freedom in these models is counted by the moduli-dependent quantity $F_1 \simeq N$ and one finds \cite{vandeHeisteeg:2022btw}
\begin{equation}
\label{LambdaspF1}
\frac{\Lambda_{\rm sp}}{M_P}\simeq \frac{1}{\sqrt{F_1}}\, .
\end{equation}
Possible ambiguities concerning the index-like nature of $F_1$ are discussed as well.

While swampland conjectures gain solidness after passing non-trivial tests in the string landscape, their most basic and universal aspects can usually be formulated independently from string theory, for example using black holes physics. 
A bottom-up black hole argument supporting the claim \eqref{LambdaspF1} is not present in \cite{vandeHeisteeg:2022btw} and in this letter we provide one.

We consider a certain class of black hole solutions in the presence of higher curvature corrections arising e.g.~from M-theory compactified on a Calabi-Yau threefold times a circle. 
Here, an M5-brane wrapping a holomorphic (very ample) divisor gives rise to an extremal black hole in the four-dimensional effective supergravity theory.
The entropy of this configuration has been computed from first principles in \cite{Maldacena:1997de} and then matched successfully with the corresponding macroscopic quantity in \cite{LopesCardoso:1998tkj}; see also \cite{Behrndt:1998eq,LopesCardoso:1999fsj,LopesCardoso:2000qm}.

Due to the presence of higher curvature corrections, the Bekenstein-Hawking contribution to the entropy receives a modification proportional to the second Chern class of the Calabi-Yau threefold. From a supergravity perspective, this correction is encoded into a contribution to the prepotential linear in the moduli, which we denote $F_1$. By studying large and small entropy limits \cite{Bonnefoy:2019nzv,Cribiori:2022cho}, we argue that such a quantity gives a lower bound to the entropy $\mathcal{S}$ of extremal black holes that can be reliably described by the effective theory,
\begin{equation}
\mathcal{S}\gtrsim F_1 \simeq N. 
\end{equation}
According to \cite{Dvali:2007hz,Dvali:2007wp,Dvali:2010vm}, this sets the species cutoff of the theory as in \eqref{LambdaspF1}.
Thus, from black hole physics we find a moduli-dependent species scale given in terms of the first order correction to the prepotential of the underlying supergravity theory.
Demanding that this scale $\Lambda_{\rm sp}$ coincides with that proposed in \cite{vandeHeisteeg:2022btw} amounts to state that the conjecture of \cite{Ooguri:2004zv} relating the black hole to the topological string partition function should hold, at least within the regime of validity of our analysis.

\section{Extremal black holes with $R^2$ corrections}

The two-derivative supergravity effective action in four dimensions can be supplemented with a variety of corrections which are in general difficult to determine, given the lack of a systematic procedure. However, one of such corrections is known explicitly together with its numerical coefficient. Microscopically, it descends from the $R^4$-term in eleven dimensions \cite{Antoniadis:1997eg}, while in string perturbation theory it is a one-loop correction, not renormalized due to anomaly cancellation \cite{Green:1997tv}. In four dimensions, it reduces to the following correction to the effective action \cite{Maldacena:1997de}
\begin{equation}
\label{Scorr}
S_{\rm corr} =\frac{1}{96 \pi} \int c_{2i}\,\, {\rm Im}\, z^i \,{\rm Tr}\, R \wedge * R\, .
\end{equation}
In $\mathcal{N}=2$ supergravity, the fields $z^i$ with $i=1,\dots,n_V$ are scalars in the vector multiplets, while $c_{2i}$ are parameters. Their microscopic expression is given in terms of the second Chern class of the Calabi-Yau threefold $M$ on which string/M-theory is compactified as
\begin{equation}
c_{2i} \equiv \int_{M} c_2(TM) \wedge \omega_i\, ,
\end{equation}
with $\omega_i \in H^2(M;\mathbb{Z})$. We are interested in the modification to the entropy of extremal black holes due to such correction.

In general, higher derivatives corrections make the use of the Bekenstein-Hawking entropy formula invalid. Instead, by means of the Wald formula, \cite{LopesCardoso:1998tkj} succeeded in calculating the macroscopic entropy of extremal black holes and matched it correctly with \cite{Maldacena:1997de}. The first step is to understand how higher derivative corrections modify the effective $\mathcal{N}=2$ supergravity action. Then, black holes solutions can be studied in such a framework. 

The approach of \cite{LopesCardoso:1998tkj} is to consider matter-coupled $\mathcal{N}=2$ supergravity with vector multiplets $X^\Lambda$, with $\Lambda=0,1,\dots, n_V$, and couple it to a background field $A$.\footnote{In general, hyper multiplets are also present in the theory. However, given that in our setup they are uncharged under the black hole electric and magnetic fields, we neglect them in the following.} In order not to break supersymmetry already at the level of the action, the background field has to be the lowest component of a (unreduced) chiral multiplet. To get the desired higher curvature correction \eqref{Scorr}, $A$ shall be given by $ A = (T^{abij}\epsilon_{ij})^2$, with Weyl and chiral weights $(w,c)=(2,-2)$. Here, $T^{abij}$ is an auxiliary field of the Weyl multiplet with frame indices $a,b=0,\dots,3$ and SU(2)$_R$ indices $i,j=1,2$. Then, one can check that the highest component of $A$ is a scalar field containing the desired higher curvature corrections and entering explicitly the Lagrangian.
The whole construction is highly non-trivial and we refer to \cite{Mohaupt:2000mj} for a comprehensive and pedagogical review.

The interactions on the vector multiplets moduli space are encoded into a holomorphic, homogeneous of degree two prepotential $F=F(X,  A)$, which can be formally expanded as
\begin{equation}
\label{Fgexp}
F(X,  A) = \sum_{g=0}^\infty F_g(X) \,  A^g.
\end{equation}
The functions $F_g(X)$ are homogeneous of degree $2-2g$ in the fields $X^\Lambda$. 
In particular, $F_0$ is the classical prepotential, while $F_1$ is the first order correction, which will give us a lower bound on the black hole entropy and thus a moduli-dependent definition of species scale. 
The reader should not confuse the first order correction $F_1$ with $\partial_1 F$; in fact, the latter is not going to be mentioned directly in what follows, so that every time we write $F_1$ we mean the correction to the prepotential.
As usual, from $F$ one can calculate all relevant quantities, such as K\"ahler potential and gauge kinetic function, by using the standard rules of special geometry.

In \cite{LopesCardoso:1998tkj}, it is proven that the only allowed $\mathcal{N}=2$ vacuum configurations with arbitrary $A$ and which are static and spherically symmetric are Minkowski and $AdS_2 \times S^2$. 
The latter arises as the near horizon geometry of extremal black hole solutions whose entropy is captured by the Wald formula \cite{Wald:1993nt}
\begin{equation}
\mathcal{S} = 2\pi \oint_{S^2} \epsilon_{ab}\epsilon_{cd} \frac{\delta \mathcal{L}}{\delta R_{abcd}}\, .
\end{equation}
Here, $\mathcal{L}$ is the supergravity Lagrangian including the curvature corrections, while $\epsilon_{ab}$, with indices $a,b=0,1$, is the bivector normal to the horizon, normalized as $\epsilon^{ab}\epsilon_{ab}=-2$.
A direct computation results in the entropy \cite{LopesCardoso:1998tkj}
\begin{equation}
\label{Sbhcorr}
\mathcal{S} = \pi\left[Z \bar Z - 256 \,{\rm Im}\, F_{A} (X, A)\right],
\end{equation}
where $Z$ is the $\mathcal{N}=2$ central charge and $F_A = \partial_A F$. 
This formula is model-independent and valid for any prepotential.
For later purposes, it will be convenient to express the Bekenstein-Hawking-like contribution $Z \bar Z$ in terms of the K\"ahler potential $K=-\log i\left(\bar X^\Lambda F_\Lambda(X,A) - X^\Lambda \bar F_{\Lambda}(\bar X, \bar A)\right)$ as 
\begin{equation}
\label{ZZb}
Z \bar Z =e^{-K(X,A)}= X^0 \bar X^0 e^{-K(z,A)}, \qquad \text{with} \qquad z^\Lambda = \frac{X^\Lambda}{X^0}.
\end{equation}

To correctly reproduce the physical entropy of extremal black holes, one has to evaluate \eqref{Sbhcorr} at the background value for $ A$ and at the attractor point for the vector multiplets scalar fields. In an appropriate frame\footnote{In \cite{LopesCardoso:1998tkj}, this frame is denoted with $(Y^\Lambda, \Upsilon)$, not to confuse it with the original frame $(X^\Lambda,  A)$ in which the action is derived. Here, in order to avoid unnecessary notation, we keep using $(X^\Lambda,  A)$.}, the former is $ A = -64$, while the latter are determined by the attractor equations \cite{Behrndt:1996jn}
\begin{align}
\label{pattr}
p^\Lambda &=i(X^\Lambda - \bar X^\Lambda),\\
\label{qattr}
q_\Lambda &=i \left(F_\Lambda(X, A) - \bar F_{\Lambda}(\bar X, \bar{ A})\right),
\end{align}
where $q_\Lambda$, $p^\Lambda$ and the electric and magnetic charges of the black hole.

As originally pointed out in \cite{Ooguri:2004zv}, the expression \eqref{Sbhcorr} for the entropy can be recast in a suggestive form. Crucially, the presence of the correction is key for this to work. 
Using the fact that $A = \bar A =-64$ for the background, we can first rewrite the entropy as
\begin{equation}
\mathcal{S} = \pi \left[ Z \bar Z + 4 {\rm Im}\, (A F_{A})\right].
\end{equation}
Next, from the homogeneity property $X^\Lambda F_\Lambda + 2 A F_A = 2F$, together with \eqref{ZZb} and the attractor equation \eqref{qattr}, we get
\begin{equation}
 \mathcal{S} = \pi \left[(X^\Lambda + \bar X^\Lambda) q_\Lambda + 4 {\rm Im}\, F\right]   .
\end{equation}
Finally, by (formally) introducing a real part $\frac{1}{2\pi}\phi^\Lambda$ for $X^\Lambda$ and also the function
\begin{equation}
\mathcal{F} (\phi, p) = 4\pi \,{\rm Im}\, F, \qquad \text{such that}\qquad \frac{\partial\mathcal{F}}{\partial \phi^\Lambda} = -q_\Lambda,
\end{equation}
we have
\begin{equation}
\mathcal{S} =  -\phi^\Lambda   \frac{\partial\mathcal{F}}{\partial \phi^\Lambda}+\mathcal{F} (\phi, p).
\end{equation}
Thus, the entropy is the Legendre transform of the functional $\mathcal{F}$ with electric chemical potential $\phi^\Lambda$.
The (mixed) black hole partition function is defined to be $\mathcal{Z}_{BH}=\exp{\mathcal{F}}$ and \cite{Ooguri:2004zv} proposed that it should coincide with the topological string partition function,
\begin{equation}
\mathcal{Z}_{BH} = |\mathcal{Z}_{\rm top}|^2.
\end{equation}
We will see that, at least in the regime of validity of our analysis, this conjecture follows if we identify the moduli-dependent species scale determined in the next section with the recent proposal of \cite{vandeHeisteeg:2022btw}.

\section{Entropy limits and the species scale}

We would like to derive a moduli-dependent expression for the species scale from a lower bound on the entropy of extremal black holes.  To this purpose, we follow the strategy of \cite{Bonnefoy:2019nzv, Cribiori:2022cho} to study large and small entropy limits  in the vector multiplets moduli space.
Even if the discussion has been so far general, we now restrict us to a particular class of models for convenience.
Notice that this is different from the class recently considered in \cite{Delgado:2022}.

We choose a prepotential of the form 
\begin{equation}
F(X,A) = -\frac{1}{6} \frac{C_{ijk} X^i X^j X^k}{X^0} + d_i \frac{X^i}{X^0} A, \qquad \text{with} \qquad d_i \equiv -\frac{1}{24}\frac{1}{64} c_{2i}.
\end{equation}
It is of the general type \eqref{Fgexp}, with a classical cubic term and a correction $F_1 = d_i \frac{X^i}{X^0}$. 
This class arises from type IIA string theory compactified on a Calabi-Yau threefold with triple intersection numbers $C_{ijk}$ or equivalently from heterotic string on $K3 \times T^2$. Whenever referring to the microscopic description, we will stick to the type IIA/M-theory frame for convenience. The classical (string frame) volume of the Calabi-Yau threefold $M$ is given by
\begin{equation}
\label{Vol}
\mathcal{V}_0 = \frac16 C_{ijk}t^it^jt^k,
\end{equation}
where $t^i ={\rm Im}\, z^i$. Validity of the supergravity description requires the restriction of the moduli space to the bulk of the K\"ahler cone $t^i >0$. 

We are interested in extremal black holes in this setup. By appropriately rotating the symplectic frame, we can choose to work with the non-vanishing charges $q\equiv-q_0$ and $p^i$ only. In supergravity they are arbitrary (quantized) parameters, but in order for the entropy to match with the microscopic calculation, we should ask for the hierarchy \cite{Maldacena:1997de,LopesCardoso:1998tkj}
\begin{equation}
\label{q>p>0}
q \gg p^i \gg 0.
\end{equation}
The first condition is necessary for the validity of the long wavelength approximation and it implies that $t^i>0$ is satisfied at the horizon. In other words, the horizon lies inside the K\"ahler cone. The second condition implies that the divisor wrapped by the M5-brane is very ample.

The vector multiplets scalar fields at the horizon are fixed by the attractor equations \eqref{pattr} and \eqref{qattr} in terms of the charges as
\begin{align}
\label{Xattr}
X^0=-\frac12 \sqrt{\frac{\frac16 C_{ijk}p^ip^jp^k+4 d_i p^i A}{q}}, \qquad X^i = -\frac i2 p^i,
\end{align}
with $A=-64$, while the entropy \eqref{Sbhcorr} results in \cite{Maldacena:1997de,LopesCardoso:1998tkj}
\begin{equation}
\mathcal{S} = 2\pi \sqrt{\frac 16 q \left(C_{ijk}p^ip^jp^k + c_{2i}p^i\right)}\, .
\end{equation}

Now, we perform an analysis similar to that of \cite{Bonnefoy:2019nzv, Cribiori:2022cho} and study large or small entropy limits induced by limits on the moduli space. A simple but universal modulus to study is the volume \eqref{Vol}. Turning off the background $A$ for one moment, the expression of the volume at the horizon would be
\begin{equation}
\label{Volhor}
\mathcal{V}_0 = \sqrt{\frac{q^3}{\frac16 C_{ijk}p^ip^jp^k}} \, .
\end{equation}
However, as it is clear from \eqref{Xattr}, a non-vanishing $A$ enters the volume in a rather complicated manner. In the following, we argue that one can still take \eqref{Volhor} as a proxy for the volume at the horizon even in the presence of a non-trivial background, at least within the supergravity regime. 
Inspired by \eqref{Fgexp}, one can set up a formal expansion $\mathcal{V}=\sum_{g=0} \mathcal{V}_g A^g$, where $\mathcal{V}_0$ is the contribution at $A=0$. 
One can then compute explicitly this series and notice that, in the regime $p^i \gg 0$, the terms $\mathcal{V}_g$ for $g>1$ are indeed small compared to $\mathcal{V}_0$. 
This justifies using \eqref{Volhor} as a proxy for the volume modulus across the charge lattice even in the presence of corrections, as long as we are within \eqref{q>p>0}. The case $p^i \to 0$ requires extra care and will be discussed more carefully in due time. For the moment, we proceed with \eqref{Volhor}.

Having identified the leading charge dependence of the volume modulus, leaving aside possible caveats for $p^i \to 0$, \cite{Bonnefoy:2019nzv, Cribiori:2022cho} suggest to look at limits of large or small charges inducing in turn large or small entropy. These should correspond to the breakdown of the effective theory at the horizon, since the black hole becomes either too small or too big. 

First, we consider the case of an isotropic limit in which all of the charges $p^i$ scale in the same way. 
Here, the aforementioned caveats for $p^i \to 0$ do not occur and $\mathcal{V}_0$ is a good control parameter. 
It is helpful to distinguish the following four possibilities:\footnote{In quantum gravity charges are expected to be quantized so that a parametrically small value is not attainable. This will not invalidate our argument, since we want to estimate a lower bound to the entropy and not to compute a precise value.}
\begin{align}
\begin{array}{lcccc}
\nonumber
\text{constant $q$} \qquad &i)\quad &p^i \to \infty &\qquad \mathcal{V}_0 \to 0 &\qquad \mathcal{S} \to \infty\\
\nonumber
 &ii)\quad &p^i \to 0 &\qquad \mathcal{V}_0 \to \infty &\qquad \mathcal{S} \to 0\\
 \nonumber
 \text{constant $p^i$} \qquad &iii)\quad &q \to \infty &\qquad \mathcal{V}_0 \to \infty & \qquad \mathcal{S} \to \infty\\
 \nonumber
 &iv)\quad &q \to 0 &\qquad \mathcal{V}_0 \to 0 &\qquad \mathcal{S} \to 0
 \end{array}
\end{align}
The limits $i)$ and $iv)$ violate the hierarchy \eqref{q>p>0} and thus we cannot draw any conclusion from supergravity, since we lose the microscopic interpretation. The limit $ii)$ leads to a small black hole which cannot be described by the supergravity approximation. Here, we expect corrections to the effective action to become important, but in this isotropic limit their effect is somehow washed out. A more careful analysis is needed and we will come back to it in a while. The limit $iii)$ is instead perfectly within the supergravity approximation \eqref{q>p>0}. The volume of the internal manifold and the entropy of the black hole both diverge. According to the general classification of \cite{Lee:2019wij}, this limit corresponds to decompactification and thus a tower of Kaluza-Klein states becoming light is predicted, invalidating the effective description. 

Now, we come back to $ii)$ but we slightly modify it in order to appreciate the effect of the higher curvature corrections we have been considering. We take a minimal non-isotropic limit in which only one of the magnetic charges, say $p^1$, is kept fixed, while the others are sent to zero. Changing the number of charges $p^i$ kept constant will not change the argument, as long as there is at least one. A limit of this kind can be engineered microscopically by choosing  $M$ to be a $K3$-fibration over $T^2$, such as the Enriques Calabi-Yau $(K3\times T^2)/\mathbb{Z}_2$, and wrapping the M5-brane on the $K3$ fiber.
In this setup, only one of the $p^i$ would be non-vanishing and thus $C_{ijk}p^ip^jp^k=0$ implying that the classical volume at the horizon diverges and the entropy is entirely supported by the correction. 

Before proceeding, let us also notice that the classical volume \eqref{Volhor} can now give misleading information. Indeed, by defining a notion of quantum volume from $K=-\log (8 \mathcal{V} X^0 \bar X^0)$, a direct computation shows that 
\begin{equation}
\mathcal{V} \to\frac12 \frac{q^{2}}{ \sqrt{\frac q6 c_{2i}p^i}},
\end{equation}
in such a modified $ii)$ limit. This will be a consistency check for our moduli-dependent expression of the species scale.

Even if we deviate minimally from the isotropic situation, the behaviour of the entropy is qualitatively different with respect to the previous case, 
\begin{equation}
\mathcal{S} \to 2\pi \sqrt{\frac q6 \, c_{2i}p^i }\, ,
\end{equation}
and one can now appreciate how the presence of the correction avoids a non-physical vanishing entropy. 
In fact, it provides a lower bound for the entropy, which can be most clearly seen by using \eqref{ZZb} to rewrite \eqref{Sbhcorr} as\footnote{We systematically set axions to zero, since these expressions are implicitly evaluated at the black hole horizon. Otherwise, one could replace $F_1 \to {\rm Im}\, F_1$ below. Furthermore, we are neglecting a factor $i$ which differs between supergravity and topological string conventions. }
\begin{equation}
\label{S=eKF1}
\mathcal{S} = \pi \left[e^{-K}+\frac16 c_{2i}\,{\rm Im}\left( \frac{X^i}{X^0}\right)\right] \gtrsim \, F_1 \simeq c_{2i}t^i .
\end{equation}
Recalling that the entropy of a black hole should be bigger than the number of light species and that the smallest black hole reliably described by the effective theory sets the species scale as the inverse of its horizon radius \cite{Dvali:2007hz,Dvali:2007wp,Dvali:2010vm}, we have 
\begin{equation}
\label{S=F1}
\mathcal{S} \gtrsim  F_1 \simeq N\simeq\frac{M_P^2}{\Lambda_{\rm sp}^2} 
\end{equation}
and  we recover \eqref{LambdaspF1} as desired. In fact, one could have arrived at \eqref{S=F1} directly from \eqref{S=eKF1} and without considering the limits $ii)$ and $iii)$. Nevertheless, we chose to discuss the four limiting cases above for pedagogical reasons and to present a technique which is general and can be applied for other purposes or in different configurations. Notice also that in the modified $ii)$ limit the exponential term in \eqref{S=eKF1} does not vanish but it competes with the second one.

A consistency check for our proposal can be performed by looking at the Kaluza-Klein scale, $M_{KK}$, which in large volume limits is related to the species scale as $\Lambda_{sp}\simeq N^{\frac16} M_{KK}$, in the case of six compact dimensions. For the setup under investigation, the Kaluza-Klein scale is estimated as $M_{KK} \simeq M_s/\mathcal{V}^\frac16 \simeq M_P g_s/\mathcal{V}^{2/3}$, giving in turn $\Lambda_{sp}\simeq M_P/\sqrt{\mathcal{V}}$. By checking the behaviour of the volume in the limits considered above we can verify explicitly when the effective theory breaks down due to towers of Kaluza-Klein states becoming massless. In the isotropic limit $iii)$, $M_{KK}$ and $\Lambda_{sp}$ vanish as $\mathcal{V}\to \infty$ and we predict a breakdown of the effective description due to an infinite tower of light states, in accordance with \cite{Lee:2019wij}. In the non-isotropic $ii)$ limit we considered next, $\mathcal{V}\to \frac12 q^{2} / \sqrt{\frac q6 c_{2i}p^i}$ does not diverge since at least one of the charges $p^i$ is kept constant. Thus, $M_{KK}$ and $\Lambda_{sp}$ do not vanish and we do not expect a tower of light states. This seems to suggest that non-isotropic limits in the moduli space could help in obtaining scale-separated flux vacua. A construction implementing this idea has been proposed recently in \cite{Cribiori:2021djm}, but whether or not these are ultimately consistent string theory vacua is an open problem. Indeed, swampland conjectures \cite{Lust:2019zwm}, see also \cite{Cribiori:2022trc, Lust:2022lfc,Montero:2022ghl}, suggest that scale-separated anti-de Sitter vacua should not be in the landscape.

Strictly speaking, what we derived is the functional form of \eqref{LambdaspF1}, but the underlying interpretation could be different from \cite{vandeHeisteeg:2022btw}, since in our case $F_1$ is a correction to the classical supergravity prepotential and not yet the genus-one partition function of the topological string. Thus, at least conceptually, the species scale that we found from our black hole argument might differ from \cite{vandeHeisteeg:2022btw}. Imposing that this should not happen and that the two scales are in fact one and the same, in accordance with \cite{Antoniadis:1993ze}, we are led to state that the conjecture of \cite{Ooguri:2004zv}, namely $\mathcal{Z}_{BH}=|\mathcal{Z}_{\rm top}|^2$, holds at least within the regime of validity of our analysis. This suggests that such a relation can play a role in the swampland program more in general.

\section{Discussion}

In this letter, we presented a black hole argument leading to a moduli-dependent definition of species scale in quantum gravity of the form \eqref{LambdaspF1}. The analysis took place on the vector multiplets moduli space of $\mathcal{N}=2$ supergravity in four dimensions, where $F_1$ contributes to the classical prepotential and encodes certain higher derivative corrections to the effective action. 

The result formally matches with the recent proposal of \cite{vandeHeisteeg:2022btw}. Asking for the physical interpretation to be the same, in order that the associated species scales coincide as well, we are led to enforce the validity of the conjecture $\mathcal{Z}_{BH}=|\mathcal{Z}_{\rm top}|^2$ relating the black hole to the topological string partition function, originally put forward in \cite{Ooguri:2004zv}. From this perspective, our argument lends further support also to the recent proposal of \cite{vandeHeisteeg:2022btw} of a moduli-dependent species scale \eqref{LambdaspF1} and, in particular, it provides a rationale behind it based on black hole physics.

More in general, our analysis connects the species scale in quantum gravity to certain higher curvature corrections to the effective action. This is a form of IR/UV mixing which is unexpected from a low energy point of view, a priori. Given that the species scale counts light degrees of freedom, one might be tempted to infer that the sign of such higher curvature corrections should be fixed in order for $F_1 \simeq N$ to be a positive number. However, care must be taken in this respect, due to the index-like nature of $F_1$.

The results here presented can be extended along various directions. For example, it would be interesting to consider black holes in asymptotically anti-de Sitter space, or to repeat a similar analysis in higher dimensions, or even to study the effects of adding a  non-vanishing temperature to the system. 
More in general, it would be interesting to investigate further the consequences of the conjecture $\mathcal{Z}_{BH}=|\mathcal{Z}_{\rm top}|^2$ within the swampland program.

\paragraph{Acknowledgments.}
The work of N.C.~is supported by the Alexander-von-Humboldt foundation.
The work of D.L.~is supported by the Origins Excellence Cluster and by the German-Israel-Project (DIP) on Holography and the Swampland.
 
\bibliography{references}  
\bibliographystyle{utphys}

\end{document}